\documentclass[prb,amsmath,amssymb,superscriptaddress,twocolumn]{revtex4}

\usepackage{amssymb}
\usepackage{graphicx}
\usepackage{color}
\usepackage{dcolumn}
\usepackage{bm}
\usepackage{multirow}

\begin{document}

\newcommand{\ir}{Sr$_2$IrO$_4$}
\newcommand{\irr}{Sr$_3$Ir$_2$O$_7$}
\newcommand{\jeff}{$J_{\rm{eff}}=1/2$}

\title{Persistent non-metallic behavior in Sr$_2$IrO$_4$ and Sr$_3$Ir$_2$O$_7$ at high pressures}

\author{D. A. Zocco}
\altaffiliation[Present address: ]{Institute for Solid State Physics (IFP), Karlsruhe Institute of Technology, D-76021 Karlsruhe, Germany}
\email[Corresponding author: ]{diego.zocco@kit.edu}
\author{J. J. Hamlin}
\altaffiliation[Present address: ]{Department of Physics, University of Florida, Gainesville, Florida 32611, USA}
\author{B. D. White}
\affiliation{Department of Physics, University of California, San Diego, La Jolla, California 92093, USA}
\author{B. J. Kim}
\affiliation{Department of Solid State Spectroscopy, Max Planck Institute for Solid State Research, D-70569 Stuttgart, Germany}
\affiliation{Randall Laboratory of Physics, University of Michigan, Ann Arbor, Michigan 48109, USA}
\author{J. R. Jeffries}
\affiliation{Condensed Matter and Materials Division, Lawrence Livermore National Laboratory, Livermore, California 94550, USA}
\author{S. T. Weir}
\affiliation{Condensed Matter and Materials Division, Lawrence Livermore National Laboratory, Livermore, California 94550, USA}
\author{Y. K. Vohra}
\affiliation{Department of Physics, University of Alabama at Birmingham, Birmingham, Alabama 35294, USA}
\author{J. W. Allen}
\affiliation{Department of Physics, Randall Laboratory, University of Michigan, Ann Arbor, Michigan 48109, USA}
\author{M. B. Maple}
\affiliation{Department of Physics, University of California, San Diego, La Jolla, California 92093, USA}

\begin{abstract}
Iridium-based 5$d$ transition-metal oxides are attractive candidates for the study of correlated electronic states due to the interplay of enhanced crystal-field, Coulomb and spin-orbit interaction energies. At ambient pressure, these conditions promote a novel \jeff\ Mott-insulating state, characterized by a gap of the order of $\sim$\,0.1\,eV. We present high-pressure electrical resistivity measurements of single crystals of \ir\ and \irr. While no indications of a pressure-induced metallic state up to 55\,GPa were found in \ir, a strong decrease of the gap energy and of the resistance of \irr\ between ambient pressure and 104\,GPa confirm that this compound is in the proximity of a metal-insulator transition.
\end{abstract}

\maketitle

\section{Introduction}

Metal-to-insulator transitions (MIT) have been widely studied in transition-metal oxides (TMOs), mainly motivated by the discoveries of high-temperature superconductivity in cuprates and colossal magnetoresistance in manganites.\cite{imada98a} As opposed to the Mott-insulating ground state found in the 3$d$-electron compound La$_{2}$CuO$_{4}$, a metallic ground state is expected to be found in iridium-based TMOs, due to the highly delocalized 5$d$ electronic orbitals of the Ir ions. However, non-metallic behavior has been found in \ir\ and \irr, which are members of the Ruddelsden-Popper series Sr$_{n+1}$Ir$_{n}$O$_{3n+1}$ ($n$ = number of IrO$_2$ layers). The unexpected insulating behavior has been attributed to the strong spin-orbit coupling (SOC\,=\,0.2\,-1\,eV) which in these compounds is comparable to the Coulomb repulsion $U$\,=\,0.5\,-\,2\,eV (SOC\,$\sim$\,0.01\,eV in 3$d$ TMOs), giving rise to a novel \jeff\ Mott-insulating ground state.\cite{kim08a,kim09a} With increasing $n$, the bandwidth $W$ associated with the 5$d$ orbitals increases and the Mott gap becomes smaller. Earlier optical spectroscopy studies\cite{moon08a} revealed an energy gap value of $\sim$\,0.1\,eV for \ir, a gap value almost equal to zero for \irr, and a metallic state in the $n\,=\,\infty$ compound, namely SrIrO$_3$. On the other hand, recent angle-resolved photoemission spectroscopy (ARPES) and scanning tunneling microscopy (STM) experiments indicate the existence of larger energy gaps, as a result of the interplay between structure, single-atom defects, SOC and correlations.\cite{okada13a,wang13a}
\begin{figure}[t]
\begin{center}
{\includegraphics[width=3.1in]{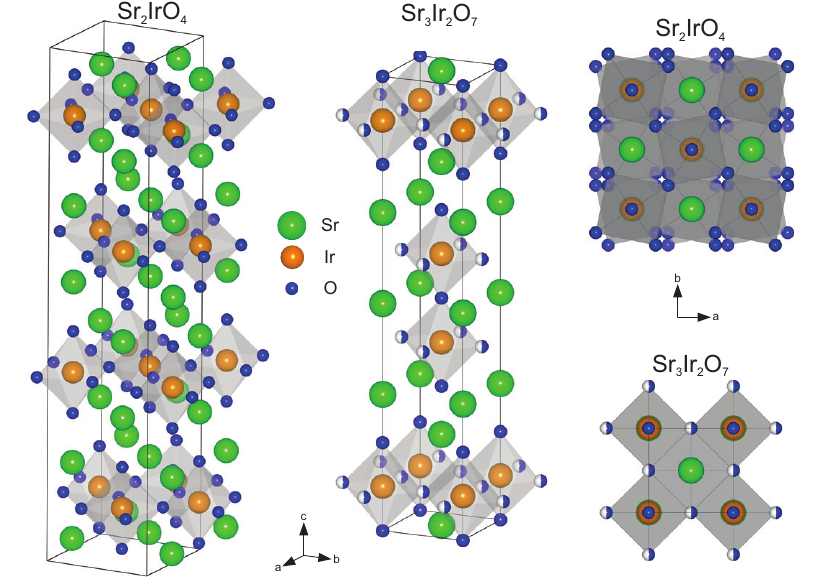}}
\end{center} \caption{(color online). Crystal structure of single-layer \ir\ and double-layer \irr. The O and Ir atoms form octahedra, which are rotated by 11$^{\circ}$ about the $c$-axis in \ir. In \irr, the IrO$_{6}$ octahedra form an undistorted $I4/mmm$ tetragonal structure.} \label{fig1}
\end{figure}

\ir\ forms in a reduced tetragonal $I4_1/acd$ structure, similar to tetragonal $I4/mmm$ but with the IrO$_{6}$ octahedra rotated by 11$^{\circ}$ about the $c$-axis, which increases the size of the unit cell.\cite{crawford94a,huang94a} It orders into an antiferromagnetic (AFM) state below $T_{\mathrm{N}}$\,=\,240\,K, with canted Ir moments in the $ab$-planes.\cite{cao98a,kim09a,junghokim12a} \irr\ forms in a tetragonal $I4/mmm$ structure, given that a rotation of the IrO$_{6}$ octahedra has not been observed.\cite{boseggia12a} The bilayer \irr\ displays long range AFM order below $T_{\mathrm{N}}$\,=\,285\,K, with collinear moments along the $c$-axis.\cite{jwkim12a,junghokim12b} It has recently been pointed out that the Ir-O-Ir bond angle controls the electronic hopping and the magnetic interaction between Ir atoms, which allows the physical properties to be tuned via the introduction of oxygen vacancies\cite{korneta10a} or chemical substitution,\cite{ge11a} or via the application of magnetic field\cite{chikara09a} and pressure\cite{haskel12a}. In this paper, we report measurements of the electrical resistivity of \ir\ and \irr\ under pressure. These experiments were motivated by the small values of the Mott-insulating gap in these compounds which one could expect to quench with pressure.

\section{Experimental details}

Single crystals of \ir\ and \irr\ were grown by means of a molten metal flux method as described in Ref.\,\onlinecite{kim09a}. Measurements of resistivity along the $ab$ plane were performed for 1\,K\,$\leq$\,$T$\,$\leq$\,300\,K under hydrostatic pressure conditions up to 1.66\,GPa employing a Be-Cu piston-cylinder hydrostatic cell (HC). A 1:1 mixture of $n$-pentane and isoamyl alcohol was used as the pressure medium, contained in a Teflon capsule. Electrical contacts were made by attaching with silver epoxy four 50\,$\mu$m Pt wires to the surface of the crystals. Pressure was determined by measuring the superconducting transition of a Sn manometer.\cite{smith69} A minimum pressure of 0.1\,GPa is needed to close the cell. Pressure gradients were inferred from the width of the superconducting transition of the manometer ($\delta P < 2\%$).

For higher pressures, a mechanically loaded diamond-anvil cell (DAC) was used. We utilized a non-beveled blank diamond with a tip of 300\,$\mu$m in diameter, and a designer diamond with a culet size of 500\,$\mu$m beveled to a tip of 250\,$\mu$m. The designer diamond contains eight deposited tungsten microprobes encapsulated in high-quality homoepitaxial diamond, for resistivity measurements.\cite{jackson03} The gasket was made from a 200\,$\mu$m thick MP35N foil pre-indented to 50\,$\mu$m, and a 100\,$\mu$m diameter hole was drilled through the center of the indentation. The sample space corresponds to the volume enclosed by the hole in the gasket and the flat tips of the two diamonds, and for these experiments, the sample was packed in this cavity, without pressure medium. Electrical contacts were obtained by pushing the leads against the sample during pressurization. Minimum pressures of 2-3\,GPa were necessary to obtain low contact resistance. The pressure was determined at room temperature, using the fluorescence spectrum of chips of ruby located inside the sample chamber, with a $\delta P$\,$\leq$15\%, inferred from the full width at half maximum (FWHM) of the fluorescence line. Pressure gradients in the sample chamber were inferred from the pressure difference between different rubies. Electrical resistance $R$ was measured with a Linear Research Inc. LR-700\,AC bridge (maximum $R$\,=\,2\,M$\Omega$) and a standard four-lead technique.

\section{Results}

\subsection{Sr$_2$IrO$_4$}
\begin{figure}[t]
\begin{center}
{\includegraphics[width=3.3in]{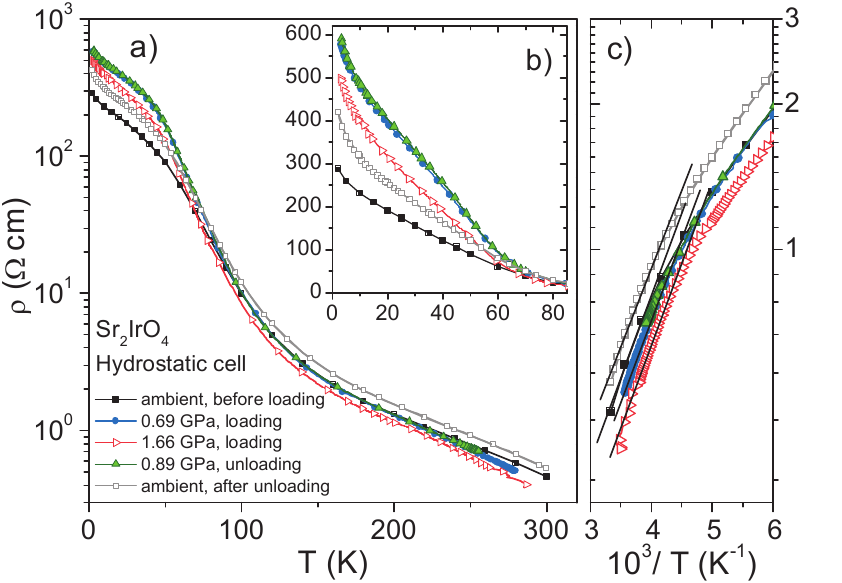}}
\end{center} \caption{(color online). Electrical resistivity $\rho$ versus temperature $T$ of \ir\ measured in the hydrostatic cell (HC) experiment, plotted in semi-logarithmic (a) and linear (b) scales. Pressure was increased from ambient pressure to 1.66\,GPa, and then released to 0.89\,GPa and back to ambient pressure. (c) Arrhenius plots for determination of the energy gap values. Solid lines represent linear fits in the region between 200\,K and 300\,K.} \label{fig2}
\end{figure}

\begin{figure}
\begin{center}
{\includegraphics[width=3.3in]{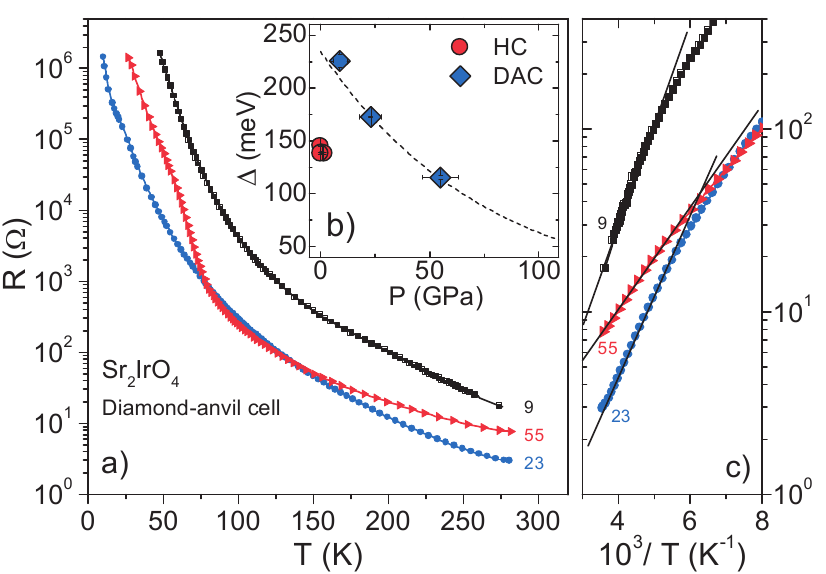}}
\end{center} \caption{(color online). (a) Electrical resistance $R$ versus temperature $T$ of \ir\ measured in the diamond-anvil cell (DAC). Numbers indicate pressure in GPa. The pressure dependence of the gap energies $\Delta$ are shown in (b), and correspond to the Arrhenius plots displayed in (c) for the DAC and in Fig.\,\ref{fig2}(c) for the HC experiments.} \label{fig3}
\end{figure}

Fig.\,\ref{fig2}(a) displays the electrical resistivity $\rho$ (logarithmic scale) versus temperature $T$ of \ir\ at ambient pressure and for pressures up to 1.66 GPa obtained in the HC experiment. All the curves show similar non-metallic behavior throughout the entire temperature range. No anomaly at $T_{\mathrm{N}}$\,$\sim$\,240\,K is observed, which is consistent with previous reports.\cite{chikara09a} From 200\,K to 300\,K, the $\rho (T)$ data exhibit activated behavior, and can be described by an exponential function $\rho(T)$\,$\propto$\,$\mathrm{exp}(\Delta/2k_{B}T)$, where $\Delta$ is the energy gap and $k_B$ is the Boltzmann constant. $\Delta$ has been estimated by plotting $\mathrm{ln}$\,$\rho (T)$ versus $1/T$ (Arrhenius plot) and fitting a straight line to the data [Fig.\,\ref{fig2}(c)]. The obtained value of $\Delta$\,$\sim$\,(140\,$\pm$\,5)\,meV is similar to the ambient-pressure value previously reported\cite{moon08a} and remains relatively unchanged within the pressure range used in the HC experiment. The value of resistivity at room temperature decreases as pressure is increased, and it recovers fairly well to the ambient pressure value ($\sim$\,0.5\,$\Omega$--cm) after unloading the clamp. At intermediate temperatures, from 100\,K to 200\,K, the data can be better described with an expression appropriate for a weakly-localized two-dimensional system, with $\mathrm{log}\,\rho$\,$\propto$\,$-T$, which can be attributed to the effects of structural disorder present in the material.\cite{kini06a}

The scattering appears weaker below 100\,K, where $\rho (T)$ can be better described by a linear temperature dependence [Fig.\,\ref{fig2}(b)]. This regime is accompanied by an irreversible pressure-dependence: at the lowest temperatures, the values of resistivity reach a maximum at intermediate pressures (0.69\,GPa-loading and 0.89\,GPa-unloading), and $\rho (T)$ does not fully recover to the initial ambient pressure curve after completely unloading the pressure clamp (for both ambient pressure curves, the sample was measured outside the pressure cell). The reduced scattering found below 100\,K, and the possible irreversible pressure dependence, might be related to the onset of magnetic order. Previously reported in-plane resistivity measurements performed at ambient pressure show a sharp feature close to this temperature, and magnetization measurements revealed evidence for magnetic order occurring below 100\,K.\cite{chikara09a,korneta10a,ge11a} In those reports, the authors suggested that this state could be originating from a gradual canting of the Ir magnetic moments due to the $T$-dependent rotation of the IrO$_{6}$ octahedra about the $c$-axis.\cite{crawford94a} This idea is supported by the strong $T$-dependence of the bending phonon modes associated with the Ir-O-Ir bond angle observed in recent optical spectroscopy experiments, which influences the magnetic exchange interaction.\cite{moon09a}

Fig.\,\ref{fig3} displays the resistance $R$ of \ir\ as a function of temperature corresponding to the diamond-anvil cell experiment.\cite{nota} \ir\ remains non-metallic up to 55 GPa, suggesting that higher pressures are necessary to drive the system to a metallic state, as has been recently pointed out by Haskel \textit{et al.}.\cite{haskel12a} In our measurements, a pronounced decrease of the overall resistance takes place when pressure is increased from 9\,GPa to 23\,GPa. On the other hand, $R(T)$ at 55\,GPa displays a flattened curvature from room temperature to $\sim$\,80\,K, followed by a sharper increase of $R(T)$. This could be related to the suppression of $T_{\mathrm{N}}$ to lower temperatures, as it is expected at these values of pressure.\cite{haskel12a} The relative increase in the room-temperature sample resistance at 55 GPa compared to that at 23 GPa is possibly due to a reduction of the thickness of the sample at this pressure.

The pressure dependence of $\Delta$ corresponding to the DAC experiment has been determined as described for the HC case. The corresponding Arrhenius plots are shown in Fig.\,\ref{fig3}(c). The resulting gap values are plotted in Fig.\,\ref{fig3}(b), showing a monotonic decrease from 225\,meV at 9\,GPa to 115\,meV at 55\,GPa. Clearly, the attained maximum pressure has not been sufficient to close the gap completely. One can notice a difference between the value of $\Delta$ at 9\,GPa and the value extracted from the HC experiments (140\,meV). Unfortunately, it was not possible to measure with the DAC at lower pressures, as described in Section 2, so we cannot distinguish between the possibility of a sudden increase of $\Delta$ between ambient pressure and 9\,GPa and the possibility of some sample dependence such as is evident in the literature\cite{moon08a,okada13a,wang13a} for these compounds. This uncertainty does not affect our main conclusions concerning the pressure dependences of the two compounds, in the discussion section below.

\subsection{Sr$_3$Ir$_2$O$_7$}
\begin{figure}[t]
\begin{center}
{\includegraphics[width=3.4in]{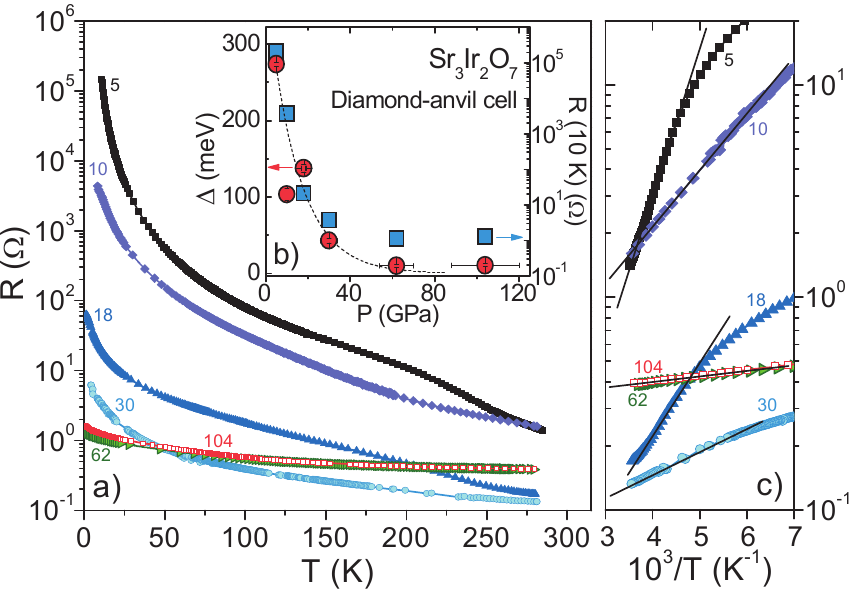}}
\end{center} \caption{(color online). (a) Electrical resistance $R$ versus temperature $T$ of \irr\ measured in two diamond-anvil cell runs. Numbers indicate pressure in GPa. Run\,1: 5, 10, 18 and 30\,GPa, Run\,2: 62 and 104\,GPa. (b) Values of $R$ measured at 10\,K (squares, right axis) and of energy gap $\Delta$ (circles, left axis) as a function of pressure, obtained from the fits to the data displayed in the (c) panel.} \label{fig4}
\end{figure}

Figure\,\ref{fig4}(a) displays $R$ (logarithmic scale) versus $T$ of \irr\ obtained in the DAC. In the first run, a maximum pressure of 30\,GPa was attained, while measurements of $R(T)$ at 62\,GPa and 104\,GPa were made during a second DAC run. A hump below room temperature develops in the 5\,GPa curve, possibly related to the onset of the magnetic order.\cite{cao02a} This feature is absent at higher pressures, probably due to the strong pressure gradients along the sample which tend to soften the anomalies observed in electrical resistivity. The values of $R$ at 10 K as a function of applied pressure are plotted in Fig.\,\ref{fig4}(b) (right axis), showing a dramatic reduction by more than five orders of magnitude. The curves at 62 and 104\,GPa are quite similar, with small temperature dependence. Under these extreme conditions, large pressure gradients develop in the sample, which are reflected in the large error bars associated with the values of pressure determined from different ruby spheres located in the sample chamber. Values of energy gap $\Delta$ have also been determined as in the \ir\ case described above. In this case, the fits to the Arrhenius plots are shown in Fig.\,\ref{fig4}(c). At 5\,GPa, the fit results in an energy gap of 270\,meV. $\Delta$ is suppressed monotonically with pressure for \irr, decreasing below 10\,meV at the highest pressures [Fig.\,\ref{fig4}(b), left axis].

\section{Discussion}

In Figs.\,\ref{fig3}(b) and \ref{fig4}(b), the gap energies from the DAC experiments are plotted as a function of pressure for \ir\ and \irr, respectively. The dashed lines correspond to an exponential suppression of the gap at a rate $\alpha$, $\Delta(P)$\,=\,$\Delta_0 \mathrm{exp}(-\alpha P)$, where $\Delta_0$ is the ambient-pressure gap. This simple empirical relation for $\Delta(P)$ can be used to roughly estimate a value of applied pressure needed to induce the metallic state. For example, tellurium is non metallic at ambient pressure and $\Delta(P)$ can also be described with this exponential formula with $\alpha$\,=\,0.63\,GPa$^{-1}$ and $\Delta_0$\,=\,323\,meV.\cite{anzin77} In this context, Te is then expected to reach an energy gap of 1\,meV at $P_{1{\rm meV}}$\,=\,9\,GPa. In fact, pure Te becomes a metal at a pressure of approximately 5\,GPa.\cite{btm64} For \ir, we obtain $\alpha$\,=\,0.014\,GPa$^{-1}$ and $P_{1{\rm meV}}$\,=\,388\,GPa, implying that much larger pressures than the ones attained in our DAC experiment are necessary to metallize this compound. For \irr, the fit results in $\alpha$\,=\,0.054\,GPa$^{-1}$ and $P_{1{\rm meV}}$\,=\,105\,GPa, in agreement with our measurements.
\begin{table}[h]
\small
\caption{\label{table1}
Parameters of the pressure dependence of the energy gap, for the expression $\Delta(P)$\,=\,$\Delta_0 \mathrm{exp}(-\alpha P)$.}
\centering
\begin{tabular}{c|c|c|c|c}
  \hline \hline
Material & Reference & $\Delta_0$ (meV) & $\alpha$ (GPa$^{-1})$ & $P_{1{\rm meV}}$ (GPa) \\
  \hline
{\multirow{2}{*}{Sr$_2$IrO$_4$}}    & this work                   & 250 & 0.014 & 388 \\

                                    & Ref.\,\onlinecite{haskel12a}  &  57.8 & 0.027 & 149 \\
  \hline
{\multirow{2}{*}{Sr$_3$Ir$_2$O$_7$}}& this work                   & 273 & 0.054 & 105 \\
                                    & Ref.\,\onlinecite{li13a}      &  2.15 & 0.15  & 5   \\
  \hline
Te                                  & Ref.\,\onlinecite{anzin77}    & 323 & 0.63  & 9   \\
  \hline \hline
\end{tabular}
\end{table}

The parameters of the exponential pressure dependence of the gap are summarized in Table\,\ref{table1}, along with the values corresponding to Te and to the data from Ref.\,\onlinecite{haskel12a} and Ref.\,\onlinecite{li13a} for \ir\ and \irr, respectively (the electrical resistivity measurements presented in Ref.\,\onlinecite{haskel12a} were performed under similar experimental conditions as in Ref.\,\onlinecite{li13a}, that is, using a powdered pressure medium). It is clear that simple exponential fits to the pressure dependences of the gap energies predict very different values of pressure needed to metallize the samples. This could be attributed to the different values of $\Delta_0$ characterizing these samples, which depend strongly on the temperature range where the fits to the Arrhenius plots are made. In this regard, we expect that the effects of thermally-activated behavior originating from a conducting energy gap would be better characterized at the higher temperature range used here rather than at lower temperatures where the conducting quasi-particles can no longer be thermally excited across the gap and other conduction mechanisms, presumably extrinsic, dominate the transport. Moreover, the energy gap has been measured already at room temperature via infrared spectroscopy, above the N\'{e}el temperature, as reported in Ref.\,\onlinecite{moon08a}. As has been pointed out in the introduction, the electronic properties of these layered iridate compounds are highly sensitive to the interplay between structure and defects, resulting in samples of quite different ambient-pressure gap energies.\cite{okada13a,wang13a} Moreover, one should expect that defects are created in the samples at the highly non-hydrostatic conditions obtained in diamond-anvil cells, which could also help to explain the large difference in the values of $\alpha$ (a factor of 2 for \ir, and a factor of 3 for \irr) between these experiments.

The possibility of a pressure-induced structural phase transition (SPT) in \ir\ and \irr\ has already been investigated. Haskel \textit{et al.}\cite{haskel12a} have not found any evidence of a pressure-induced SPT in \ir\ up to 25 GPa (helium pressure medium, x-ray synchrotron measurements). Similar structural refinement under pressure has not yet been performed on \irr. Li \textit{et al.}\cite{li13a} observed a sharp decrease of the electrical resistivity at 13.2 GPa for this compound, indicative of a SPT. In measurements of electrical resistivity under pressure, a first-order change in structure often results in an abrupt jump in the resistivity as pressure is changed at a fixed temperature. It is essential to be able to discard other effects that could cause the abrupt change in resistivity, such as changes in the thickness of the sample or changes in the geometry of the electrical leads. In this regard, we have not observed any indications of a pressure-induced SPT in our measurements.

For the \irr\ compound, the extremely weak temperature dependence of the resistance measured at 62 and 104\,GPa suggests that this system could be on the verge of a MIT transition. Recent high-pressure experiments suggest that it could actually take place at much lower pressures.\cite{li13a} It is surprising that a pressure increment by 42\,GPa, from 62\,GPa to 104\,GPa, did not significantly affect the transport properties of \irr. A possible explanation could be that pressure did increase locally in the regions where the ruby chips where located, but not in the region of the sample located in between the voltage leads. Such large pressure gradients could lead to an inhomogeneous metallic and insulating material, which could explain the lack of an observed pressure-induced metallic state.

One should also ask whether higher pressures would actually drive the system into a metallic state. In Ref.\,\onlinecite{korneta10a}, it has been shown that the introduction of very small amounts of oxygen vacancies into single crystals of Sr$_2$IrO$_{4-\delta}$ led to a MIT for $\delta$\,$\sim$\,0.04. This small amount of doping, however, increased the Ir-O-Ir bond angle $\theta$ by less than 1$^{\circ}$, far less than the previous estimation of $\Delta\theta\sim 13^{\circ}$ necessary to close the $\sim$100\,meV Mott gap.\cite{moon09a} This suggests that structural changes induced by pressure could be insufficient to transform the system to a metallic state. A study of the relation between the Ir-O-Ir bond angle and the transport properties at high pressures might shed light on this subject.

In summary, we have measured the electrical resistivity of \ir\ and \irr\ under externally applied pressures, up to 55\,GPa and 104\,GPa, respectively. In both cases, no definitive signatures of a pressure-induced metallic state have been found up to the maximum pressures achieved. For \irr, however, the strong suppression of the energy gap and of the resistance measured at 10\,K, confirm that this compound is in the proximity of a metal-insulator transition.

\section{Acknowledgments}

High-pressure research at UC San Diego was supported by the National Nuclear Security Administration (NNSA) under the Stewardship Science Academic Alliance program through the US Department of Energy (DOE) grant number DE-52-09NA29459. Physical properties characterization at ambient pressure was supported by DOE Grant DE-FG02-04-ER46105. LLNL is operated by Lawrence Livermore National Security, LLC, for the DOE-NNSA, under Contract No. DE-AC52-07NA27344. Y.K.V.\ acknowledges support from the DOE-NNSA Grant No. DE-NA0002014. This work was supported at UM by the U S National Science Foundation under grant number DMR-07-04480. BJK acknowledges the Institute for Complex Adaptive Matter for a travel grant that enabled a visit to UC San Diego and thereby led to the conception of this work.


\begin{thebibliography}{xx}
\bibitem{nota} The DAC experiment performed in the Sr$_2$IrO$_4$ sample was stopped
  at 55\,GPa. The hole in the gasket, observed through the upper diamond during
  pressurization, was growing to a size which could lead to a sudden release of
  the pressure and a consequent damage of the anvils. For the same reason, the
  DAC experiment for Sr$_3$Ir$_2$O$_7$ was stopped at 104\,GPa.
\end{thebibliography}
\end{document}